\documentclass[12pt, a4paper, twoside]{fec}
\usepackage{graphicx,epsf}
\usepackage{mathrsfs}
\usepackage{color}
\usepackage{dcolumn}
\usepackage{multicol}
\usepackage{amsmath}
\usepackage{amssymb}
\begin{document}

\title{Nonlinear saturation of   toroidal Alfv\'en eigenmodes via nonlinear mode couplings}


\author[1]{Zhiyong  Qiu}
\author[1,2]{Liu Chen}
\author[3,1]{Fulvio Zonca}
\author[4]{Wei Chen}
\affil[1]{Institute for Fusion Theory and Simulation, Zhejiang University, Hangzhou, P.R.C.}
\affil[2]{Department of Physics and Astronomy,  University of California, Irvine CA, U.S.A.}
\affil[3]{ENEA, Fusion and Nuclear Safety Department, C. R. Frascati, Frascati (Roma), Italy}
\affil[4]{Southwestern Institute of Physics, P.O. Box 432 Chengdu 610041, PeopleÕs Republic of China}

\doi

\emailID{zqiu@zju.edu.cn}

\PaperNumber{TH/P2-6}

\maketitle

\linespread{1}  \normalsize
\begin{abstract}
{
Gyrokinetic theory of nonlinear mode coupling as a mechanism for toroidal Alfv\'en eigenmode (TAE) saturation   in the fusion plasma related parameter regime is presented, including 1) parametric decay of TAE into lower kinetic TAE (LKTAE) and geodesic acoustic mode (GAM), and 2) enhanced TAE coupling to shear Alfv\'en wave (SAW) continuum via ion induced scattering.  Our theory shows that, for TAE saturation in the parameter range of practical interest, several processes with comparable scattering cross sections can be equally important.
}
\end{abstract}
\linespread{1}  \normalsize

\section{Introduction}\label{sec:intro}

Shear Alfv\'en wave (SAW) instabilities, manifested as various Alfv\'en eigenmodes (AE) due to equilibrium magnetic geometry and/or plasma nonuniformity in tokamaks,  such as toroidal Alfv\'en eigenmode (TAE) \cite{CZChengAP1985},  are expected to play important roles in burning plasmas with energetic particle (EP) contributing significantly to the total energy density  \cite{LChenRMP2016}.   TAE can be strongly resonantly driven by EPs,   and lead to EP transport and degrade overall plasma confinement, as reviewed and discussed in Ref. \cite{LChenRMP2016}. The transport rate is related to TAE amplitude, and thus, understanding the nonlinear dynamics of TAE including saturation,  is important for accessing the properties of burning plasmas in future reactors.  In this paper, gyrokinetic theory of nonlinear mode coupling as a potential mechanism for TAE saturation  in the burning plasma related parameter regime is presented, including 1) parametric decay of TAE into lower kinetic TAE (LKTAE) and geodesic acoustic mode (GAM), and 2) enhanced TAE coupling to SAW  continuum via ion induced scattering.  The consequences on EP transport and thermal plasma heating, are also discussed.
Our theory shows that, for TAE saturation in the parameter range of practical interest, several processes with comparable scattering cross sections can be equally important \cite{LChenPRL2012,ZQiuPRL2018}.

\section{Parametric decay of TAE into LKTAE and GAM}\label{sec:TAE_GAM_LKTAE}

In this section, we consider the decay of a TAE into a GAM and LKTAE in the continuum as one possible mechanism for TAE nonlinear  saturation. The process considered here, besides the apparent effects on TAE nonlinear saturation, may lead  to plasma heating via ion Landau damping of the secondary GAMs. The generated GAM, as the finite frequency component of zonal structures, may interact with, e.g., drift wave turbulences,  leading to improved confinement.

To investigate the nonlinear interactions among  the pump TAE ($\omega_0$, $\mathbf{k}_0$),   GAM ($\omega_G$, $\mathbf{k}_G$) and   LKTAE ($\omega_L$, $\mathbf{k}_L$), the scalar potential $\delta \phi$ and parallel vector potential $\delta A_{\parallel}$ are adopt as the field variables. One then has, $\delta\phi=\delta\phi_0+\delta\phi_G+\delta\phi_L$, with the subscripts $0$, $G$ and $S$ denoting pump TAE, GAM and LKTAE, respectively.
Furthermore, $\delta\psi\equiv \omega\delta A_{\parallel}/(ck_{\parallel})$ is taken as an alternative variable, and one has $\delta\psi=\delta\phi$ in the ideal MHD limit.
For TAE and lower KTAE with high toroidal mode numbers in burning plasmas, we adopt the well-known ballooning-mode decomposition   in the $(r,\theta,\phi)$ field-aligned flux coordinates
\begin{eqnarray}
\delta\phi_0&=&A_0e^{i(n\phi-m_0\theta-\omega_0t)}\sum_j e^{-ij\theta}\Phi_0(x-j)+c.c.,\nonumber\\
\delta\phi_L&=&A_Le^{i(n\phi-m_0\theta-\omega_0t)}e^{-i(\int \hat{k}_Gdr-\omega_Gt)} \sum_je^{-ij\theta}\Phi_L(x-j)+c.c..\nonumber
\end{eqnarray}
Here, $(m=m_0 +j, n)$ are the poloidal and toroidal mode
numbers, $m_0$ is the reference value of $m$,
$nq(r_0) =m_0$, $q(r)$ is the safety factor, $x=nq-m_0 =
nq'(r-r_0)$,  $\hat{k}_G$ is the radial envelope due to GAM modulation and  $\hat{k}_G\equiv nq'\theta_k$  in the  ballooning representation, $\Phi$ is the fine radial structure associated with $k_{\parallel}$ and magnetic shear, and $A$ is the envelope amplitude.

While for the secondarily generated GAM, we assume it is predominantly electrostatic, and has both the usual meso- scale structure  and an additional fine-scale radial structure due to the radially localized structure of the pump TAE
\begin{eqnarray}
\delta\phi_G&=&A_Ge^{i(\int \hat{k}_Gdr-\omega_Gt)}\sum_j \Phi_G(x-j)+c.c..\nonumber
\end{eqnarray}
Here, $\Phi_G$ is the fine scale structure of GAM \cite{ZQiuNF2016}, and the summation over $j$ is the summation over the radial positions where the pump TAE poloidal harmonics are  localized. As a result,
$\mathbf{k}_G=\mathbf{\hat{k}}_G-i\partial_r\ln\Phi_G \hat{\mathbf{e}}_r$,
and one  typically has $|\partial_r \ln\Phi_G|\gg |\hat{k}_G|$.

The governing equations describing the nonlinear interactions among TAE, LKTAE and GAM, can then be derived from quasi-neutrality condition
and nonlinear gyrokinetic vorticity equation, and the particle responses are derived from nonlinear gyrokinetic equation,  noting the   $|k_{\parallel}v_{t,e}|\gg |\omega|\simeq |V_A/(2qR_0)|\gg |k_{\parallel}v_{t,i}|\gg|\omega_{d,e}|, |\omega_{d,e}|$ orderings for TAE/LKTAE while  the $\omega_G\sim |v_{i,t}/R_0|\gg |\omega_{tr,i}|, |\omega_d|$  orderings for GAM.

The nonlinear vorticity equation of LKTAE is derived as
\begin{eqnarray}
\mathscr{E}_L\delta\phi_L=i(c/B)k_{G}k_{\theta,0}\left[(\Gamma_0-\Gamma_G)/\omega_L +(1-\Gamma_L)\sigma_0/(\sigma_L\omega_0)\right]\delta\phi_{G^*}\delta\phi_{0}.\label{eq:LKTAE}
\end{eqnarray}
Here, $\mathscr{E}_L\equiv (1-\Gamma_L)-k^2_{\parallel}V^2_A\sigma_Lb_L/\omega^2_L$ is the WKB dispersion relation of LKTAE, with $\sigma_T=1+\tau-\tau\Gamma_T\equiv \delta\psi_T/\delta\phi_T$  describing the deviation from ideal MHD condition and generation of parallel electric field due to kinetic effects,  subscript $T$ denoting modes in the TAE frequency range, $\tau\equiv T_e/T_i$,  $\Gamma_k\equiv \langle J^2_k F_0/n_0 \rangle$ and  $k_{\theta,0}=-m/r$ is the poloidal mode number of $\Omega_0(\omega_0, \mathbf{k}_0)$.
Noting that $\omega_L=\omega_0-\omega_G$, the nonlinear coupling coefficient of equation (\ref{eq:LKTAE}), recovers  that of equation (10) (taking only ``-" for lower sideband) of Ref. \cite{FZoncaEPL2015} for KAW lower sideband  generation by pump KAW beating with  finite frequency convective cell, when only the electro-static convective cell generation is considered, and assuming $|\omega_G|\ll|\omega_0|$ in the small $\beta$ limit.

The LKTAE eigenmode dispersion relation  can be derived, noting that  $V^2_A\propto (1-2(r/R_0+\Delta')\cos\theta)$   due to toroidicity  \cite{FZoncaPoP1996}  with $\Delta'$ being the Shafranov shift.  We then obtain
\begin{eqnarray}
\hat{\mathscr{E}}_LA_L=i (c/B_0)k_{\theta,0}\hat{\alpha}_LA_{G^*}A_0,\label{eq:KTAE_global}
\end{eqnarray}
with
$\hat{\mathscr{E}}_L\equiv \int dr |\Phi_L|^2  \mathscr{E}_L$ being the LKTAE eigenmode dispersion relation, and  the contribution of toroidal geometry and mode structure on the nonlinear coupling fully accounted for in
$\hat{\alpha}_L \equiv  \int dr |\Phi_0|^2|\Phi_L|^2  k_G  \left((\Gamma_0-\Gamma_G)/\omega_L+(1-\Gamma_L)\sigma_0/(\sigma_L\omega_0)\right)$.

The nonlinear GAM  equation in the electrostatic limit can be determined from the nonlinear vorticity equation. Noting that LKTAE could be heavily radiative damped, it is necessary to keep both its linear and nonlinear particle responses   while deriving the nonlinear GAM equation.  After some tedious but straightforward algebra, one obtains
\begin{eqnarray}
&& \mathscr{E}_{G^*} \delta\phi_{G^*}=i(c/(B\omega_G))k_Gk_{\theta,0} \left[\Gamma_0-\Gamma_L-(b_s-b_0) k^2_{\parallel}V^2_A \sigma_{0^*}\sigma_L/(\omega_0\omega_L)\right]\delta\phi_{0^*}\delta\phi_L. \label{eq:GAM_local}
\end{eqnarray}
The two terms on the right hand side of equation (\ref{eq:GAM_local}) are the generalized Reynolds and Maxwell stresses, valid for arbitrary $k_{\perp}\rho_i$.
Here, $\mathscr{E}_G$ is the linear dispersion function of GAM,   defined as \cite{FZoncaEPL2008}
$\mathscr{E}_G\equiv \left\langle (1-J^2_G) F_0/n_0 \right\rangle- T_i \sum_s\overline{ \langle  q_s J_G\omega_d\delta H^L_{G} \rangle} /(n_0e^2\omega\overline{\delta\phi}_{G})$.
Taking $\Phi_{G^*}\equiv\Phi_{0^*}\Phi_L$ as the fast varying component \cite{ZQiuNF2016} of GAM, one then have the eigenmode dispersion relation of GAM:
\begin{eqnarray}
 \mathscr{E}_{G^*} A_{G^*}=i(c/(B\omega_G))k_{\theta,0}\hat{\alpha}_GA_{0^*}A_L. \label{eq:GAM_global}
\end{eqnarray}
Here,
$
\hat{\alpha}_G \equiv  \int \Phi_{0^*}\Phi_S k_G\left[\Gamma_0-\Gamma_S-(b_s-b_0) k^2_{\parallel}V^2_A \sigma_{0^*}\sigma_S/(\omega_0\omega_S)\right]dr/\left(\int \Phi_{0^*}\Phi_Sdr\right)$.

The nonlinear dispersion relation can then be derived from equations (\ref{eq:GAM_global}) and (\ref{eq:KTAE_global})
\begin{eqnarray}
\hat{\mathscr{E}}_L \mathscr{E}_{G^*} =-\left(ck_{\theta,0}/B_0\right)^2(\hat{\alpha}_G\hat{\alpha}_L/\omega_G) |A_0|^2. \label{eq:NL_DR_general}
\end{eqnarray}
In the long wavelength limit,  equation (\ref{eq:NL_DR_general}) recovers  equation (17) of Ref. \cite{ZQiuEPL2013}, where a pump TAE decaying into a GAM and a TAE lower sideband is discussed.

Note that, LKTAE could be heavily radiative damped, depending on plasma parameters. In Ref. \cite{ZQiuPRL2018}, the parametric decay process is investigated assuming LKTAE is a weakly damped normal mode.  In the following, we consider another limit that LKTAE is a heavily damped quasi-mode.
Noting that,
$\mathscr{E}_{G^*}=-2i b_G  \left(\gamma+\gamma_G\right)/\omega_G $,
with $\gamma_G$ being the collisionless damping rate of GAM \cite{ZQiuPPCF2009},  and that   LKTAE could be heavily radiative damped with $\hat{\mathscr{E}}_{L,I}$  comparable to $\hat{\mathscr{E}}_{L,R}$ and $\hat{\mathscr{E}}_{L,R}$ and $\hat{\mathscr{E}}_{SL,I}$ are respectively  the Hermitian and anti-Hermitian parts of $\hat{\mathscr{E}}_L$, one then has the GAM growth rate
\begin{eqnarray}
\gamma&=& -\gamma_G -   \left(ck_{\theta,0}/B_0|A_0|\right)^2(\hat{\alpha}_G\hat{\alpha}_L/(2\hat{b}_G|\hat{\mathscr{E}}_L|^2))(i\hat{\mathscr{E}}_{L,R}+ \hat{\mathscr{E}}_{L,I}).
\end{eqnarray}

So GAM can be excited through this proposed mechanism,  given that,  first, $\hat{\alpha}_G\hat{\alpha}_L\hat{\mathscr{E}}_{L,I}<0$ for effective drive; and second
\begin{eqnarray}
 -   \left(ck_{\theta,0}/B_0|A_0|\right)^2(\hat{\alpha}_G\hat{\alpha}_L/(2\hat{b}_G|\hat{\mathscr{E}}_L|^2))  \hat{\mathscr{E}}_{L,I} >\gamma_G
\end{eqnarray}
which determines the threshold condition on TAE amplitude to overcome the   GAM collisionless damping.
These two conditions, generally, requires numerical solution since they have complex dependence on the mode structures, and thus, equilibrium geometry. However, analytical estimations can be made in the simplified limits, e.g., for $b_S\ll 1$ which is though, not the general case for KTAE.  Noting that for $|b|\ll1$,  $\Gamma_k(b_k)\simeq 1-b_k-3 b^2_k/4$ and $\sigma_k\simeq 1+\tau (b_k+3 b^2_k/4)$,  one then has
$
\hat{\alpha}_G \simeq   k_G (\hat{b}_L-\hat{b}_0)\left(1- \omega^2_A/(4\omega_0\omega_L)\right)<0$, and
$\hat{\alpha}_L \simeq  (k^2_G/\omega_L)\left(\hat{b}_G-\hat{b}_0+ \hat{b}_S (1-\omega_G/\omega_0)(1+\tau\hat{b}_0) /(1+\tau\hat{b}_L)\right)>0$.
At the same time,  $\hat{\mathscr{E}}_{L,I}>0$ for heavily damped LKTAE. Thus, $\hat{\alpha}_G\hat{\alpha}_L\hat{\mathscr{E}}_{L,I}<0$ is satisfied, and radiative damping of LKTAE indeed leads to the effective drive of GAM.

Noting that $|k_{r,0}|\sim O(nq'/\epsilon_0)$,  $|k_{r,L}|\simeq O( (\epsilon_0\rho^2_i/(n^2q'^2))^{-1/4})\gg |k_{r,0}|$,  $|k_{G}|= |k_{r,0}+k_{r,L}|\simeq|k_{r,L}|$,  and that  $|\delta B_r|\simeq |k_{\theta}\delta A_{\parallel}|$, the threshold condition for the nonlinear process can be estimated as, in the $|\hat{b}_L|\ll1$ limit:
$\left| \delta B_r/B_0 \right|^2\sim (\gamma_G/\omega_G)(k^2_{\parallel,0}/\hat{b}_L)(|\hat{\mathscr{E}}_L|^2/\hat{\mathscr{E}}_{L,I})$. The eigenmode dispersion relation of LKTAE, can be written as \cite{FZoncaPoP1996,FZoncaPoP2014b}
$
\hat{\mathscr{E}}_L\equiv -(\pi k^2_{\theta}\rho^2_i\omega^2_A/(2^{2\hat{\xi}+1}\Gamma^2_L(\hat{\xi}+1/2)\omega^2_L))\left[ 2\sqrt{2}\Gamma(\hat{\xi}+1/2)/(\hat{\alpha} \Gamma(\hat{\xi}))+\delta W_f\right]$, with $\Gamma(\hat{\xi})$ and $\Gamma(\hat{\xi}+1/2)$ being gamma-functions, $\hat{\xi}\equiv 1/4-\Gamma_+\Gamma_-/(4\sqrt{\Gamma_-\hat{s}^2\hat{\rho}^2_K})$, $\Gamma_{\pm}\equiv \omega^2_L/\omega^2_A\pm\epsilon_0\omega^2_L/\omega^2_A-1/4$ determining the lower and upper accumulational point of toroidicity induced gap,  $\omega^2_A\equiv V^2_A/(q^2R^2_0)$,   $\hat{\alpha}^2=1/(2\sqrt{\Gamma_-\hat{s}^2\hat{\rho}^2_K})$ and $\hat{\rho}^2_K\equiv (k^2_{\theta}\rho^2_i/2)(3/4+T_e/T_i)$ denoting the kinetic effects. The threshold condition can then be roughly estimated as $|\delta B_r/B_0|^2\sim 10^{-9}$, comparable to other channels for TAE nonlinear saturation  via wave-wave coupling, e.g., ZFZS generation \cite{LChenPRL2012,ZQiuEPL2013}.   The consequences of the present decay process, considering LKTAE being a heavily damped quasi-mode,  on TAE nonlinear saturation and plasma heating, can be analyzed following closely that of Ref. \cite{ZQiuPRL2018}, and will be reported in a future publication.

\section{Gyrokinetic theory of TAE nonlinear saturation via ion Compton scattering}

Another channel for TAE nonlinear saturation is ion Compton scattering, first investigated in
 Ref. \cite{TSHahmPRL1995},   considered that there exists many TAEs   in the system,  located at different radial positions with their frequency  slightly different   by local parameters. TAEs are characterized by $|k_{\parallel}|\simeq 1/(2qR_0)$. Thus,  two counter-propagating TAEs  with radially overlapped mode structures  can couple  and generate   ion   sound-wave like mode with much lower frequency, and  $|k_{\parallel}|\simeq 1/(qR_0)$; which,  in turn, induces the TAE spectral transfer of fluctuation energy towards lower frequency TAEs. The wave energy is, eventually, absorbed by linearly stable   lower frequency TAEs with stronger coupling to SAW continuum.  The theory of Ref. \cite{TSHahmPRL1995}    assumed the long wavelength MHD limit with $\omega/\Omega_{ci}\gg k^2_{\perp}\rho^2_i$,  where the nonlinear couplings occur through the parallel  ponderomotive force  induced by $\mathbf{b}\cdot\delta \mathbf{J}\times \delta\mathbf{B}$ nonlinearity. For future burning plasmas of interest, however, the parameters usually fall in the short wavelength   $k^2_{\perp}\rho^2_{it}>\omega/\Omega_{ci}$ range \cite{LChenRMP2016}, and the perpendicular coupling due to Reynolds and Maxwell stresses  may dominate, leading to much lower TAE saturation level  than the prediction of Ref. \cite{TSHahmPRL1995} and consequently, much lower EP transport.

In this section, we generalize the theory of Ref. \cite{TSHahmPRL1995} to fusion plasma  relevant short wavelength regime using nonlinear gyrokinetic theory.   The analysis, following closely that of Ref. \cite{TSHahmPRL1995},  though the   nonlinear coupling cross-section is much bigger, due to the contribution of Reynolds and Maxwell stresses in the short wavelength limit.

To investigate the nonlinear TAE spectrum evolution, we adopt the standard nonlinear perturbation theory, and consider a
test TAE $\Omega_0=(\omega_0$, $\mathbf{k}_0)$   interacting with a background TAE    $\Omega_1=(\omega_1$, $\mathbf{k}_1)$ and generating an ion sound wave like mode  $\Omega_S=(\omega_S$, $\mathbf{k}_S)$. The subscripts $0$, $1$ and $S$ denote test TAE, background TAE   and ion sound mode, respectively.     $\Omega_0=\Omega_1+\Omega_S$ is adopted as the frequency/wavenumber matching condition. For effective spectrum transfer by nonlinear ion Landau damping,  we have $|\omega_S|\sim O(v_{it}/(qR_0))$, i.e., the ion sound mode frequency comparable to thermal ion transit frequency.
Therefore,    $\Omega_0$ and $\Omega_1$ are counter-propagating TAEs,  with $\omega_0\simeq\omega_1$ and $k_{\parallel,0}\simeq -k_{\parallel,1}$.
Low-$\beta$ (plasma to magnetic pressure ratio) is also assumed, i.e., $\beta\ll\epsilon^2$, such that the nonlinearly generated  sideband is also  a TAE in the toroidicity induced gap, as is the case of Ref. \cite{TSHahmPRL1995}. In the high-$\beta$ limit with $\beta>\epsilon^2$, however, the pump TAE decay into a LKTAE in the continuum in one single step, and the TAE saturation process becomes very different. The TAE decay in high-$\beta$ limit will be investigated in another publication \cite{ZQiuNF2018}.

\subsection{Nonlinear parametric instability}

The nonlinear generation of ion sound mode due to $\Omega_0$ and $\Omega_1$ beating, is derived from
\begin{eqnarray}
\mathscr{E}_S\delta\phi_S=i(\hat{\Lambda}/\omega_0)\beta_1\delta\phi_0\delta\phi_{1^*}, \label{eq:NL_quasimode}
\end{eqnarray}
where
$\hat{\Lambda}\equiv (c/B_0)\mathbf{\hat{b}}\cdot\mathbf{k}_0\times\mathbf{k}_{1^*}$, $\mathscr{E}_S\equiv 1+\tau+\tau\Gamma_S\xi_SZ(\xi_S)$ is the linear dispersion function of $\Omega_S$, with  $\xi_S\equiv\omega_S/(k_{\parallel,S}v_{it})$ and $Z(\xi_S)$ is the  plasma dispersion function.
Furthermore,  $\beta_1\equiv \sigma_0\sigma_{1}+\tau\hat{F}_1\left(1+\xi_SZ(\xi_S)\right)$,  with $\hat{F_1}\equiv\langle J_0J_1J_SF_0/n_0\rangle$, $\sigma_k\equiv 1+\tau-\tau\Gamma_k$.

Since  $\Omega_S$ could be heavily  ion Landau damped, one needs to include both its linear and nonlinear responses while deriving the nonlinear particle responses to $\Omega_0$. Substituting it into the quasi-neutrality condition,   one has
\begin{eqnarray}
\delta\psi_0=\left(\sigma_0+\sigma^{(2)}_0\right)\delta\phi_0+D_0\delta\phi_1\delta\phi_{S},\label{eq:NL_QN_test}
\end{eqnarray}
in which,
$\sigma^{(2)}_0\equiv\hat{\Lambda}^2\left[-\sigma^2_1\sigma_0+\tau\hat{F}_2\left(1+\xi_SZ(\xi_S)\right)\right] |\delta\phi_1|^2/\omega^2_0$ and
$D_0\equiv i\hat{\Lambda}\tau\hat{F}_1\left[1+\xi_S Z(\xi_S)\right]/\omega_0$.
The   nonlinear eigenmode equation of $\Omega_0$, can be derived from vorticity equation as
\begin{eqnarray}
&&\left(\mathscr{E}_0+ \mathscr{E}^{NL}_0\right)\delta\phi_0=- \left(D_2 \omega^2_0/\hat{b}_0 +   k^2_{\parallel,0}V^2_A D_0\right)\delta\phi_1\delta\phi_{S}. \label{eq:NL_test_DR_WKB}
\end{eqnarray}
Here, $\mathscr{E}_0\equiv\epsilon_T(\Omega_0)$ is the WKB linear dispersion relation of   $\Omega_0$, with
$\mathscr{E}_T\equiv k^2_{\parallel,T}V^2_A \sigma_T-(1-\Gamma_T)\omega^2_T/\hat{b}_T$, and $\mathscr{E}^{NL}_0\equiv - \alpha^{(2)}_0/\hat{b}_0+k^2_{\parallel,0}V^2_A\sigma^{(2)}_0$
with
 $\alpha^{(2)}_0 = \hat{\Lambda}^2  \left(\hat{F}_2-\hat{F}_1\right)\left(1+\xi_SZ(\xi_S)\right)  |\delta\phi_1|^2$ and
$D_2=-i \hat{\Lambda}\left[\hat{F}_1(1+\xi_SZ(\xi_S))-\Gamma_S\xi_SZ(\xi_S)-\Gamma_1\right]/{\omega_0}$.
Note that, the TAE WKB dispersion relation defined here, has a $1\omega^2_T/\hat{b}$ coefficient compared to that of Sec.\ref{sec:TAE_GAM_LKTAE}. $\sigma^{(2)}$ and $\alpha^{(2)}$ correspond, respectively, to the contribution of  nonlinear particle response to $\Omega_S$ on  ideal MHD constraint breaking  and Reynolds stress.

Substituting equation (\ref{eq:NL_quasimode}) into (\ref{eq:NL_test_DR_WKB}), we obtain
\begin{eqnarray}
\left( \mathscr{E}_0+ \mathscr{E}^{NL}_0\right)\delta\phi_0=  -(\hat{\Lambda}^2\beta_1\beta_2/(\hat{b}_0\tau\mathscr{E}_S))|\delta\phi_1|^2\delta\phi_0,\label{eq:NL_test_DR_3}
\end{eqnarray}
with $\beta_2\equiv\beta_1/\sigma_0-\mathscr{E}_S$.
Equation (\ref{eq:NL_test_DR_3}) describes the nonlinear evolution of the test TAE $\Omega_0$ due to the nonlinear interactions with $\Omega_1$. Since  ion Compton scattering related to $\Omega_S$  ion Landau damping may  play an important role  for TAE saturation, we write the coefficients explicitly as  functions of  $\mathscr{E}_S$, i.e.,
$\mathscr{E}^{NL}_0=- (\hat{\Lambda}^2/\hat{b}_0)|\delta\phi_1|^2\left(\hat{G}_1+\hat{G}_2\mathscr{E}_S\right)$ with
$\hat{G}_1=(1-\Gamma_0)\sigma^2_1- \sigma_S\hat{G}_2$ and
$\hat{G}_2=\left(\hat{F}_2-\hat{F}_1-(1-\Gamma_0)\tau\hat{F}_2/\sigma_0\right)/(\tau\Gamma_S)$.  On the other hand,
$\beta_1\beta_2/(\tau\mathscr{E}_S)=\hat{H}_1+\hat{H}_2\mathscr{E}_S+ \hat{H}_3/\mathscr{E}_S$ with
$\hat{H}_1= \left(\sigma_0\sigma_1-\hat{F}_1\sigma_S/\Gamma_S\right)\left(2\hat{F}_1/\Gamma_S-\sigma_0\right)/(\tau\sigma_0)$,
$\hat{H}_2= \hat{F}_1 \left(\hat{F}_1/\Gamma_S-\sigma_0\right)/(\tau\sigma_0\Gamma_S)$, and $\hat{H}_3 =  \left(\sigma_0\sigma_1-\hat{F}_1\sigma_S/\Gamma_S\right)^2/(\tau\sigma_0)$.
The nonlinear $\Omega_0$ eigenmode dispersion relation, can then be derived, by multiplying both sides of equation (\ref{eq:NL_test_DR_3}) with $\Phi^*_0$, noting  that $\mathscr{E}_S$ varies much slower than $|\Phi_0|^2$ and $|\Phi_1|^2$ in radial direction, and integrating over the radial domain. One then has
\begin{eqnarray}
\mathscr{E}_S\left(\hat{\mathscr{E}}_0-\Delta_0 |A_1|^2-\chi_0\mathscr{E}_S |A_1|^2 \right)A_0 =-\hat{C}_0|A_1|^2A_0,\label{eq:NL_test_DR_4}
\end{eqnarray}
in which  $\hat{\mathscr{E}}_0$ is the linear TAE eigenmode dispersion relation, defined as $\hat{\mathscr{E}}_0=\int |\Phi_0|^2 \mathscr{E}_0 dr$. The coefficients, $\Delta_0$, $\chi_0$ and $\hat{C}_0$, corresponding respectively to nonlinear frequency shift, ion Compton scattering and shielded-ion scattering, are given as
$\Delta_0=\langle\langle \hat{\Lambda}^2(\hat{G}_1-\hat{H}_1)/\hat{b}_0\rangle\rangle$,  $\chi_0= \langle\langle\hat{\Lambda}^2(\hat{G}_2-\hat{H}_2)/\hat{b}_0\rangle\rangle$,  $\hat{C}_0=\langle\langle\hat{\Lambda}^2\hat{H}_3/\hat{b}_0\rangle\rangle$, with $\langle\langle\cdots\rangle\rangle\equiv\int (\cdots) |\Phi_0|^2|\Phi_1|^2dr$ accounting for the contribution of TAE fine scale mode structures.  $\chi_0$ can be further simplified, and yields
$\chi_0=\langle\langle \hat{\Lambda}^2\left(\hat{F}_2-\hat{F}^2_1/\Gamma_S\right)/(\tau\hat{b}_0\sigma_0\Gamma_S)\rangle\rangle$, which is positive definite.

Equation (\ref{eq:NL_test_DR_4}) can be considered as the equation describing nonlinear parametric decay of a pump TAE ($\Omega_1$) into  TAE ($\Omega_0$) and    ion sound mode ($\Omega_S$) daughter waves, which can be solved for the condition of $\Omega_1$ spontaneous decay. Note  that $\Omega_S$ could be heavily ion Landau damped, depending on plasma parameter regime such as  $\tau$,    two parameter regimes with distinct decay mechanisms shall be discussed separately.

For   weakly damped $\Omega_S$  due to, e.g., $\tau\gg1$, both $\Omega_S$ and $\Omega_0$ are normal modes of the system,  and the parametric dispersion relation is given as
\begin{eqnarray}
(\gamma+\gamma_S)(\gamma+\gamma_0)=  \hat{C}_0|A_1|^2/(\partial_{\omega_S}\epsilon_{S,R}\partial_{\omega_0}\hat{\mathscr{E}}_{0,R}),
\end{eqnarray}
with $\gamma_S$ and $\gamma_0$ being, respectively, the damping rates of $\Omega_S$ and $\Omega_0$, and the subscript ``R" denoting real part.

On the other hand, for $\tau\sim O(1)$,  $\Omega_S$ is heavily ion Landau damped, and becomes a quasi-mode. One then obtains, from the imaginary part of equation (\ref{eq:NL_test_DR_4}),
\begin{eqnarray}
\gamma+\gamma_0= |A_1|^2\left( \hat{C}_0/|\mathscr{E}_S|^2+\chi_0 \right)\mathscr{E}_{S,i}/(\partial_{\omega_0}\hat{\mathscr{E}}_{0,R}),
\end{eqnarray}
and the parametric instability $\gamma>0$  requires $\omega_1>\omega_0$; i.e., the parametric decay can spontaneously  happen only when the pump TAE frequency is higher than that of the sideband, and the parametric decay process leads to, power transfer from higher to lower frequency part of the spectrum, that is, downward spectrum cascading \cite{TSHahmPRL1995}.

\subsection{TAE spectral transfer and saturation due to nonlinear ion scattering}\label{sec:spectrum_evolution}

Summation over all the background TAEs within the strong interaction region,  i.e., counter-propagating and radially overlapping with $\Omega_k$, and the frequency difference $|\omega_k-\omega_{k_1}|$ comparable with ion transit frequency ($|v_i/(qR_0)|$), denoting TAEs with their eigenfrequencies, i.e., $I_{k}\rightarrow I_{\omega}$,    the summation over ``$k_1$" can be replaced by integration over ``$\omega$",  given many background TAEs within the strong interaction range with $\Omega_{\omega}$ (continuum limit). One then has, the wave-kinetic equation describing the TAE spectral transfer:
\begin{eqnarray}
\left(\partial_t-2\gamma_{L}(\omega)\right)I_{\omega}=\frac{2}{\partial_{\omega}\mathscr{E}_{\omega,R}}\int^{\omega_M}_{\omega_L} d\omega' V(\omega,\omega')I_{\omega'}I_{\omega}, \label{eq:spectrum_temporal_evolution}
\end{eqnarray}
with  $I_{\omega}\equiv |\nabla A_{\omega}|^2$, $V(\omega,\omega')\equiv  \left(\hat{C}/|\mathscr{E}_S|^2+\chi_0\right)\mathscr{E}_{S,i}/\hat{b}_{\omega'}$,  $\omega_M$ being the highest frequency for TAE to be linearly unstable, $\omega_L$ being the lowest frequency for $I_{\omega_L}>0$. For TAEs excited in the toroidicity induced gap to minimize continuum damping, one has $\omega_M-\omega_L\simeq O(\epsilon)\omega_{T}$, comparable with the TAE gap width.

The nonlinear saturation condition can then be obtained from $\partial_tI_{\omega}=0$:
\begin{eqnarray}
\gamma_L(\omega)=-(\partial_{\omega}\mathscr{E}_{\omega,R})^{-1}\int^{\omega_M}_{\omega_L} d\omega' V(\omega,\omega')I_{\omega'}.
\end{eqnarray}
Noting that  $I_{\omega'}$ varies in $\omega'$ much slower than $V(\omega,\omega')$, and   $I_{\omega'}\simeq I_{\omega}-\omega_S\partial_{\omega}I_{\omega}$,  we have
\begin{eqnarray}
\gamma_L(\omega)
&=& \left(U_0I_{\omega}-U_1\partial_{\omega}I_{\omega}\right)/(2\omega).\label{eq:spectrum_evolution}
\end{eqnarray}
Here, $U_0 \equiv  \int^{\omega-\omega_L}_{\omega-\omega_M} d\omega_S V(\omega_S)$ and
$U_1 \equiv  \int^{\omega-\omega_L}_{\omega-\omega_M} d\omega_S \omega_S V(\omega_S)$.
Noting that, for the ion Compton scattering process to be important, one requires $\omega_M-\omega_L\gg v_{it}/(qR_0)$ and that $V(\omega_S)\propto\mathscr{E}_{S,i}$ is an odd function of $\omega_S$  varying on the scale of $v_{it}/(qR_0)$,  one then obtain
\begin{eqnarray}
I_{\omega}&\simeq& I_M(\omega_M)+\frac{1}{\overline{U_1}}\int^{\omega_M}_{\omega}\omega\gamma_L(\omega) d\omega,\label{eq:u1_bar}
\end{eqnarray}
with $I_M(\omega_M)\equiv I_{\omega}(\omega=\omega_M)$, $\overline{U}_1 \simeq  \pi^{3/2}\left( C/|\mathscr{E}_S|^2 +\chi_0\right)k^2_{\parallel,S}v^2_{it}/( 2\hat{b}_{\omega})$.
The value of $I_{\omega}$ at $\omega_M$,  $I_M(\omega_M)$, on the other hand, can be determined  noting that for $|\omega-\omega_M|\ll |k_{\parallel,S}v_{it}|$ and replacing the lower and upper integral limits of $U_0$ and $U_0$  by, $0$ and $\infty$, and one has
$U_0(\omega_M) \simeq  \overline{U_1}/(k_{\parallel,S}v_{it})$ and
$U_1(\omega_M) \simeq \overline{U_1}/2$.
$I_M(\omega_M)$ can then be derived from equation (\ref{eq:spectrum_evolution}), noting that $|U_0I_{\omega}/(U_1\partial_{\omega}I_{\omega})|\sim |(\omega_M-\omega_L)/(k_{\parallel,S}v_{it})|\gg1$, and one has
\begin{eqnarray}
I_{\omega}=\frac{2k_{\parallel,S}v_{it}\omega_M\gamma_L(\omega_M)}{\overline{U_1}}+\frac{2}{\overline{U_1}}\int^{\omega_M}_{\omega}\omega\gamma_L(\omega) d\omega.\end{eqnarray}

The overall TAE intensity at saturation, can be derived by integrating the intensity over the fluctuation population zone, and we have
\begin{eqnarray}
I_S\equiv \int^{\omega_M}_{\omega_L} I_{\omega} d\omega\simeq \frac{\overline{\gamma_L}}{\overline{U_1}}\omega^3_T\left(1-\frac{\omega_M}{\omega_L}\right)^2.\label{eq:overall_intensity}
\end{eqnarray}
In deriving equation (\ref{eq:overall_intensity}), we replaced the TAE linear growth rate $\gamma_L$ with its spectrum averaged value, $\gamma_L(\omega)\simeq\overline{\gamma_L}$, which is validated by the fact that, for burning plasma relevant parameter regimes,  a broad  TAE spectrum with comparable linear growth rate can be driven unstable. The saturation level of the magnetic fluctuations, can then be derived as
\begin{eqnarray}
|\delta B_r|^2\sim \frac{\epsilon^2\epsilon^2_{eff}}{2\pi^{3/2}}\frac{\omega_T\overline{\gamma}_Lk^2_r}{(\hat{C}/|\mathscr{E}_S|^2|+\chi_0)\Omega^2_{ci}}
\end{eqnarray}
with $\epsilon_{eff}\equiv 1-\omega_M/\omega_L\sim O(\epsilon)$ following Ref. \cite{TSHahmPRL1995}, and $|k_{\theta,T}/k_{r,T}|\simeq \epsilon$ for TAEs is assumed. This suggests that, for TAE saturation in the parameter range of practical interest, several processes with comparable scattering cross sections may be equally important \cite{LChenRMP2016,ZQiuNF2017}.

The obtained TAE saturation level and spectrum, can then be applied to derive the ion heating rate from ion nonlinear Landau damping \cite{TSHahmPST2015} and the EP transport coefficient, which will be reported in a future publication \cite{ZQiuNF2018}.

\section*{Acknowledgements}

This work is supported by
  National Key R\&D Program of China under Grant  No.  2017YFE0301900,   EUROfusion Consortium
under grant agreement No. 633053 and  US DoE GRANT.

\providecommand{\newblock}{}


\end{document}